\documentclass[aps,prb,twocolumn,superscriptaddress,showpacs]{revtex4}
\usepackage{graphicx}
\begin{document}
\title{\bf Magnetic phase diagram of a quasi-one-dimensional quantum spin system}

\author{A.~A.~Zvyagin}
\affiliation{Institut f\"ur Festk\"orperphysik, Technische
Universit\"{a}t Dresden, 01069 Dresden, Germany}
\affiliation{B.I.~Verkin Institute for Low Temperature Physics and
Engineering of the National Academy of Sciences of Ukraine, Kharkov,
61103, Ukraine}

\date{\today}

\begin{abstract}
We propose an analytical ansatz, using which the ordering
temperature of a quasi-one-dimensional (quasi-1D) antiferromagnetic
(AF) system (weakly coupled quantum spin-1/2 chains) in the presence
of the external magnetic field is calculated. The field dependence
of the critical exponents for correlation functions of 1D subsystems
plays a very important role. It determines the region of possible
re-entrant phase transition, governed by the field. It is shown how
the quantum critical point between two phases of the 1D subsystem,
caused by spin-frustrating next-nearest neighbor (NNN) and
multi-spin ring-like exchanges, affects the field dependence of the
ordering temperature. Our results qualitatively agree with the
features, observed in experiments on quasi-1D AF systems.
\end{abstract}

\maketitle

\section{Introduction}

The progress in preparation of quantum spin substances with well
defined 1D subsystems has motivated the interest in studies of them
during last years. Another reason for the investigation of
properties of quasi-1D spin systems is the relatively rare
possibility of comparison experimental data with results of exact
theories for many-body models. According to the Mermin-Wagner
theorem, \cite{MW} totally 1D spin systems with isotropic spin-spin
interactions cannot have a magnetic ordering at nonzero
temperatures. However, for quasi-1D spin systems, which 1D
subsystems have gapless spectrum of low-lying excitations, the
magnetic susceptibility, specific heat, and muon spin relaxation
often manifest peculiarities, characteristic for phase transitions
to magnetically ordered states. The critical temperature of the
ordering of a quasi-1D AF Heisenberg spin system was first
calculated in Ref.~\onlinecite{S} in the absence of the external
magnetic field. However, for quasi-1D spin systems, in which (the
largest) exchange constants along the distinguished direction are
relatively small ($\sim 1-20$~K), the transition to the ordered
state can be governed by the external magnetic field. Nowadays in
low-temperature experiments high values of the magnetic field (about
20~T for stationary fields and about 60~T for pulse fields) can be
used. Therefore, for many quasi-1D spin systems it is possible to
investigate experimentally how the external magnetic field affects
the N\'eel ordering, i.e. to determine the H-T phase diagram.

This is why, the objection of the present study is to construct an
analytical theory (convenient for comparison with experiments),
which has to show how the external magnetic field affects the
magnetic ordering in a system of weakly coupled AF Heisenberg spin
chains. To calculate the ordering temperature of the quasi-1D system
we use the mean field approximation for a weak inter-chain
interactions. In this approach 1D subsystems are considered as
clusters (of infinite size), for the description of which we can use
non-perturbative results. \cite{Zb} As a result, we propose a
relatively simple analytical ansatz for the magnetic field
dependence for the N\'eel temperature of a quasi-1D Heisenberg AF
system.

\section{Mean-field approximation}

Consider a three-dimensional spin-1/2 system with AF interactions
between spins, which form a hyper-cubic lattice. In the quasi-1D
situation the Hamiltonians of the 1D subsystems are Heisenberg
Hamiltonians of AF spin chains
\begin{equation}
{\cal H}_{1D} = J\sum_n ({\bf S}_n\cdot {\bf S}_{n+1}) -H\sum_n
S_n^z \ , \label{H1}
\end{equation}
where $J >0$ is the AF exchange coupling between nearest neighbor
spins in the chain, $H= g\mu_B B$, $B$ is the magnetic field, $g$ is
the $g$-factor of magnetic ions, and $\mu_B$ is Bohr's magneton.
Denote by $J'\ll J$ the weak inter-chain coupling between spins
belonging to different 1D subsystems of the quasi-1D system. If the
system is AF-ordered, we can write the magnetization of the $n$-th
site of the system as
\begin{equation}
{\bf M}_n = M{\bf e}_z + (-1)^n m_N {\bf e}_x \ , \label{M}
\end{equation}
where ${\bf e}_{x,z}$ are the unit vectors in the $x$- or $z$
directions, $M$ is the average magnetization, and $m_N$ is the
staggered magnetization in the direction, perpendicular to the
external field (the order parameter in the considered case). The
inter-chain interaction can be taken into account in the mean field
approximation. In that approximation in the AF phase we write the
Hamiltonian of the total system as
\begin{eqnarray}
&&{\cal H}_{mf} = {\cal H}_{1D} + zJ'M\sum_n S^z_n \nonumber \\
&&- h_N\sum_n (-1)^n S_n^x + {\rm const} \ , \label{meanf}
\end{eqnarray}
where $h_N =zJ'm_N$, and $z$ is the coordination number. The order
parameter $m_N$ (or $h_N$) has to be determined self-consistently.
The self-consistency equation reads $m_N = M_{N}(H,h_{N},T)$, where
$M_{N}(H,h_{N},T)$ is the magnetization per site of the 1D subsystem
in the effective field $H - MzJ'$ at the temperature $T$. In other
words, the susceptibility of the quasi-1D system can be written in
the mean field approximation as
\begin{equation}
\chi_{q1D} = {\chi_N\over 1-zJ'\chi_N} \ , \label{chiq1D}
\end{equation}
and the ordering takes place at the values of the temperature and
the field, at which the denominator becomes zero. Then the
transition temperature to the ordered state has to be determined
from the equation
\begin{eqnarray}
&&1 = zJ'\chi_{N} \ , \nonumber \\
&&\chi_{N} = ( \partial M_{N}(H,h_{N},T)/
\partial h_{N})_{h_{N} \to 0} \ .
\label{selfcons}
\end{eqnarray}
Notice that $\chi_N$ is exponentially small for the situation with
gapped low-energy eigenstates of the spin chain. It takes place,
e.g., for the Heisenberg spin chain for $H > H_s$ in the ground
state ($H_s =2J$ is the critical value of the magnetic field, at
which the spin chain undergoes a quantum phase transition to the
spin-saturated phase). In that case weak couplings $J'$ cannot yield
a magnetically ordered state of a quasi-1D system. Therefore, in the
following we consider only the case with gapless low-energy
eigenstates of the spin chain.

\section{Susceptibility of the one-dimensional subsystem}

The non-uniform static susceptibility of the 1D subsystem at low
temperatures can be written as
\begin{equation}
\chi_{\alpha}(q,T) = -i \sum_n \int dt e^{-iqn} \Theta (t)\langle
[S^{\alpha}(n,t), S^{\alpha}(0,0)]\rangle_{T} \ , \label{susc}
\end{equation}
where $q$ is the wave vector, $\alpha =x,y,z$, and $\langle ...
\rangle_T$ denotes the thermal average at the temperature $T$.
Asymptotic behavior of correlation functions for an integrable spin
chain for the gapless case can be obtained in the conformal field
theory limit, \cite{Zb} and it is possible to write the staggered
part of the correlation function for the transverse to the magnetic
field components in the ground state (related to $\chi_N$) as
\begin{equation}
\langle S^x_n(t) S^x_0(0) \rangle \approx (-1)^n {C\over
[n^2-(vt)^2]^{\eta/2}} + \dots \ , \label{corr}
\end{equation}
where $v$ is the Fermi velocity of low-energy excitations, $\eta =
1/2Z^2$ is the correlation function exponent, $Z$ is the dressed
charge of low-lying excitations (low-energy eigenstates), and $C$ is
a non-universal constant. \cite{LZ} Asymptotic behavior can be
extended for weak nonzero temperatures using the conformal mapping
$(n \pm vt) \to (v/\pi T)\sinh [\pi T(n \pm vt)/v]$. Then, we can
calculate susceptibilities for $q=\pi$ (we use the main
approximation), Fourier transforming of the conformal mapping of
Eq.~(\ref{corr}) at low temperatures as
\begin{equation}
\chi_N = {C\over v} \left({2\pi T\over v}\right)^{2-\eta}
B^2\left({\eta\over4},{2-\eta\over 2}\right) \ ,  \label{chiN}
\end{equation}
where $B(x,y)=\Gamma(x)\Gamma(y)/\Gamma(x+y)$ is the Euler's beta
function. Finally, we obtain the expression for the N\'eel
temperature below which the magnetic ordering takes place
\begin{equation}
T_N \approx {v\over 2\pi} \biggl[ C {zJ'\over v}\sin \left({\pi
\eta\over 2}\right) B^2\biggl({\eta\over 4}, {2-\eta\over
2}\biggr)\biggr]^{1\over 2-\eta} \ . \label{TN}
\end{equation}

\section{Bethe ansatz approach}

Fermi velocity and the critical exponent $\eta$ can be calculated
exactly using the Bethe ansatz. \cite{Zb} In the ground state phase
with gapless low-energy eigenstates at $H < H_s$, we can write
$Z=\xi(A)$, $v=\varepsilon(A)/2\pi \sigma(A)$, where $\xi(x)$ and
$\sigma(x)$ and $\varepsilon(A)$ are determined from the solution of
the Fredholm integral equations of the second kind
\begin{eqnarray}
 \nonumber \\
&&\sigma(x) +{1\over 2\pi}\int_{-A}^{A} dy {4\sigma(y)\over
(x-y)^2+4} = {1\over \pi(1+x^2)} \ , \nonumber \\
&&\rho(x) +{1\over 2\pi}\int_{-A}^{A} dy {4\rho(y)\over (x-y)^2+4} =
-{4x\over 2\pi[1+(x-A)^2]} \ , \nonumber \\
&&\xi(x) + {1\over 2\pi}\int_{-A}^{A} dy {4\xi(y)\over (x-y)^2+4} =1
\ , \label{int}
\end{eqnarray}
and
\begin{equation}
\varepsilon(A) ={4A\over (1+A^2)^2} +
\int_{-A}^{A} dx\rho(x) \left(H -{2J\over x^2+1} \right) \ . \label{en}
\end{equation}
The boundaries of integrations are related to the value of the
magnetic field ($0\le H \le H_s$ for the phase with gapless
excitations, while for $H > H_s$ the spin chain is in the
spin-saturated phase with gapped excitations) via $H=2\pi
J\sigma(A)/\xi(A)$. Equations (\ref{int}) can be solved analytically
only in some limiting cases, and numerically in other cases. In the
absence of interactions between $z$-components of neighboring spins
(so-called XY model, the Hamiltonian of which can be exactly mapped
to the one of the non-interacting fermion model using the
Jordan-Wigner transformation), the dressed charge is equal to 1. It
is also equal to unity for the isotropic Heisenberg chain at
$H=H_s$, where $A=0$. In the absence of the magnetic field we have
$A=\infty$, and the solution of integral equations can be obtained
by the Fourier transformation, which yields $v=\pi J/2$, and
$Z=1/\sqrt{2}$. The numerical solution for intermediate values of
$A$ shows that the dressed charge as a function of $H$ grows from
$1/\sqrt{2}$ to 1 for $0\le H \le H_s$, see, e.g.,
Ref.~\onlinecite{Zb}, i.e. $\eta$ decreases from 1 to 1/2 in this
domain of field values. Similarly, the velocity of low-energy
excitations decreases with the growth of the field from $\pi J/2$ to
zero in the domain $0 \le H \le H_s$. Numerical solution for the
N\'eel temperature was given, e.g., in Ref.~\onlinecite{WH}.

\section{Simple analytic ansatz}

It is not convenient, however, from the viewpoint of application of
the results for comparisons with experimental data to use numerical
solutions. This is why, we propose the simple ansatz for the
magnetic field behavior of the the velocity $v$ and correlation
function exponent $\eta$, valid in the interval $0 \le H \le H_s$:
\begin{eqnarray}
&&v = {\pi J\over 2} \sqrt{[1-(H/H_s)][1-(H/H_s)+(2H/\pi J)]} \ ,
\nonumber \\
&&\eta = {\sqrt{4f^2 - 3H^2}\over 2f} \ , \ f=\pi J\left(1 - {H\over
H_s}\right) + H\ . \label{ans}
\end{eqnarray}
The non-universal constant is equal to 0.18 at $H=0$ and near the
saturation it behaves approximately as $C\sim 0.18\sqrt{1-2M}$,
where $M$ is the average spin moment per site, cf.
Ref.~\onlinecite{LZ}, which leads to the field dependence $C\approx
0.18[2(H_s-H)/2\pi H_s]^{1/4}$ in the vicinity of the critical
saturation point $H_s$. Our ansatz is exact at the points $H=0$ and
$H=H_s$ and in the vicinity of $H=H_s$. The main deviations of our
ansatz from exact results take place for intermediate field values,
between zero and $H_s$. In the above expression for the N\'eel
temperature (and in the ones for the velocity and the critical
exponent) we did not take into account logarithmic corrections,
which exist for the characteristics of the isotropic Heisenberg AF
spin-1/2 chain near $H=0$, see, e.g., Ref.~\onlinecite{Zb}. Those
corrections can be taken into account, which yield for the
susceptibility \cite{BETG}
\begin{equation}
\chi_N \to \chi_N {\sqrt{\ln (24.27 J/T)}\over (2\pi)^{3/2}} \ .
\label{renormchi}
\end{equation}
Then for the N\'eel temperature we can modify our Eq.~(\ref{TN}) as
(cf. \cite{LZ}):
\begin{equation}
C \to (2\pi)^{-7/4}\sqrt{\sqrt{2(H_s-H)/H_s}\ln([48.54 \pi J/v)]} \
. \label{ren}
\end{equation}
Equations (\ref{TN}), (\ref{ans}), and (\ref{ren}) are the main
result of our work.

\section{Results for the magnetic phase diagram}

\begin{figure}
\begin{center}
\includegraphics[width=0.35\textwidth]{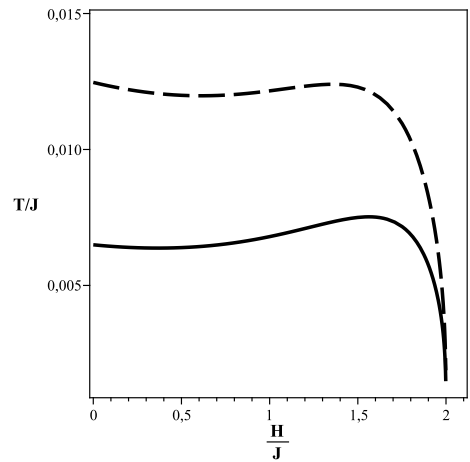}
\end{center}
\caption{The $H-T$ phase diagram of the quasi-1D Heisenberg spin-1/2
system with AF interactions between nearest neighbors (the value of
the spin saturation field $H_s=2J$). The dashed line represents the
result without logarithmic corrections, while the solid one is
related to the case with taken into account logarithmic
corrections.} \label{fig1}
\end{figure}

In Fig.~\ref{fig1} we present the N\'eel temperature of a quasi-1D
spin-1/2 AF system for $z=4$ and $J'=0.1J$ as a function of the
external magnetic field $H$, i.e. the $H-T$ phase diagram of the
system (our results qualitatively agree with the results of
numerical calculations for Bethe ansatz equations, see, e.g.,
Ref.~\onlinecite{WH}). The quasi-1D system is in the magnetically
ordered state in the interval of fields and temperatures, limited by
the line of the second order phase transition.
Logarithmic corrections do not change the qualitative behavior of
the N\'eel temperature as a function of the field. However, the
values of the critical temperatures become smaller due to
logarithmic corrections. The N\'eel temperature as a function of the
external field first grows, and then goes to zero at the critical
field $H=H_s$. Hence, there exists a (narrow) interval of
temperatures, at which one can observe a re-entrant phase
transition. In this domain of temperatures, if we enlarge the value
of the field, a quasi-1D spin system is first in the paramagnetic
short-range phase. Then the system undergoes a phase transition to
the magnetically ordered phase, and then, for  larger values of the
field, it returns to the paramagnetic phase. It would be interesting
to observe such a re-entrant phase transition in real quasi-1D spin
systems. Such a behavior is the consequence of the field dependence
of the critical exponent. If the critical exponent does not depend
on the magnetic field (e.g., in the XY chain), the N\'eel
temperature as a function of the field only decreases (following the
field dependence of the velocity). Exact calculation of the N\'eel
temperature for quasi-1D spin system with the Dzyaloshinskii-Moriya
(DM) interaction \cite{BETG} also revealed similar to Fig.~1
behavior, i.e. the maximum in the field dependence of the critical
temperature. Our calculations for spin systems with DM interactions
or with the ``easy-plane'' magnetic anisotropy show that such
magnetically anisotropic interactions produce the reduction of the
maximum in the field dependence of the critical temperature. Notice
that the critical temperature for weakly coupled XY spin chains is
higher than for Heisenberg chains due to field-independent exponent.
Also, if the symmetry of the lattice of the total system is lower
than hyper-cubic, the ordering can take place not at $q=\pi$, but
for some values of $q$, which depend on the lattice structure and
relativistic interactions (such a case can be analyzed in the random
phase approximation). In that case the N\'eel temperature also
becomes smaller than in the hyper-cubic situation, cf.
Ref.~\onlinecite{BETG}.

\section{Effect of next-nearest neighbor and multi-spin exchange
couplings}

In real quasi-1D spin systems additional intra-chain exchange
interactions between NNN spins exist very often. \cite{expfr}. To
take into account such interactions we can consider the modified 1D
Hamiltonian
\begin{equation}
{\cal H}_{NNN} = J_1\sum_n ({\bf S}_n\cdot {\bf S}_{n+1}) +J_2\sum_n
({\bf S}_n\cdot {\bf S}_{n+2}) -H\sum_n S_n^z \ , \label{H2}
\end{equation}
where $J_2$ is the exchange integral for next-nearest neighbor
couplings. For $J_2 >0$ such a spin chain reveals a spin
frustration. Unfortunately, for this Hamiltonian an exact solution
cannot be obtained analytically for any values of $J_{1,2}$.
Nevertheless, approximate bosonization studies and numerical
calculations suggest that for $J_2 > 0.24...J_1$ the spin gap is
opened for low-energy excitations. \cite{ON} For the above mentioned
reasons a system of weakly coupled chains with gapped excitations
cannot be ordered magnetically. However, as follows from
Ref.~\onlinecite{expfr}, despite the fact that for most of studied
compounds exchange constants satisfy the condition $J_2 >
0.24...J_1$, the spin gap was not confirmed experimentally. To
describe theoretically quasi-1D spin systems with spin frustration
due to intra-chain interactions without spin gap and with a weak
inter-chain coupling, we consider another model, the Hamiltonian of
which is ${\cal H}_{NNN}$ with additional terms, describing
multi-spin ring-like interactions. The advantage of that model is
its exact integrability: The model permits an exact Bethe ansatz
solution. We do not state, naturally, that the model describes all
features of the experiments. \cite{expfr} However, many properties
of the model are similar to what was observed in
Ref.~\onlinecite{expfr}, at least, for this model low-energy
eigenstates are gapless. Hence, from this viewpoint, it
qualitatively agrees with the data of experiments, \cite{expfr}
unlike the model with the Hamiltonian ${\cal H}_{NNN}$. Multi-spin
ring exchange interactions are often present in oxides of transition
metals, where a direct exchange between magnetic ions is
complimented by a super-exchange between magnetic ions via
nonmagnetic ones. \cite{exp} The modified Hamiltonian of such 1D
subsystem has the form
\begin{eqnarray}
&&{\cal H}_{mod} = {\cal H}_{NNN} + J_4\sum_n ( ({\bf S}_{n-1}{\bf
S}_{n+1}) ({\bf S}_{n}{\bf S}_{n+2}) \nonumber \\
&&- ({\bf S}_{n-1}{\bf S}_{n+2}) ({\bf S}_{n}{\bf S}_{n+1}) ) \ .
\label{H4} \end{eqnarray} Notice that multi-spin interactions are
less relevant from the renormalization group viewpoint than two-spin
interactions. Quantum properties of the model can be seen from the
exact solution, \cite{MT} for the parametrization of coupling
constants $J_1=J(1-y)$, $J_2 =Jy/2$, $J_4=2Jy$ for any $J$ and $y$
(in what follows we consider $J>0$, $y \ge 0$). This exactly
solvable model, while being formally less realistic than the model
with the Hamiltonian ${\cal H}_{NNN}$, reveals features, more
similar to the properties of experimentally studied quasi-1D systems
with spin frustration. \cite{expfr} For $y=0$ the model describes
the Heisenberg spin-1/2 chain. The ground state of the model depends
on values of the parameter $y$ and an external magnetic field.
\cite{MT} At $T=0$ for large values of the magnetic field the model
is in the spin-saturated phase, divided from other phases by the
line of the second order quantum phase transition. For low values of
$y$ and $H$ the model is in the phase, which properties are similar
to the phase of the Heisenberg spin-1/2 chain in a weak magnetic
field (Luttinger liquid). \cite{MT} The model is in this phase for
$y < y_{cr} = 4/\pi^2$ at $H=0$ and for $y < y_{cr}(H)$ for nonzero
fields. The point $y_{cr}$ is the quantum critical one. For $y >
y_{cr}$ the model is in an incommensurate phase with nonzero
spontaneous magnetization at $H=0$. Last two phases are divided from
each other by the line of the second order quantum phase transition.
Quantum phase transitions can be observed in the temperature
behavior of thermodynamic characteristics of the model, like the
magnetic susceptibility and the specific heat, that were also
calculated exactly. \cite{MT} This model permits us to know, how NNN
interactions (together with the multi-spin ones), which cause the
quantum phase transition, can modify the $H-T$ phase diagram,
obtained above for the case of only nearest neighbor couplings in
chains. In this situation the expressions for the velocity and the
critical exponent can be modified using the substitution $\pi J \to
\pi J(1- x)$ (cf. Ref.~\onlinecite{MT}), where $x=y/y_{cr}$. We
concentrate on the case $y < y_{cr}$ ($0 \le x \le 1$) for the 1D
subsystem, which has Luttinger liquid properties. In this case a
quasi-1D spin system undergoes the transition to the AF ordered
state. \cite{ZD} Ordering in the incommensurate phase was studied in
Ref.~\onlinecite{ZD}. The results of our analysis are presented in
Fig.~2. The ordered phase is inside the region, limited by the
surface, at which the second-order phase transition takes place.
\begin{figure}
\begin{center}
\includegraphics[width=0.5\textwidth]{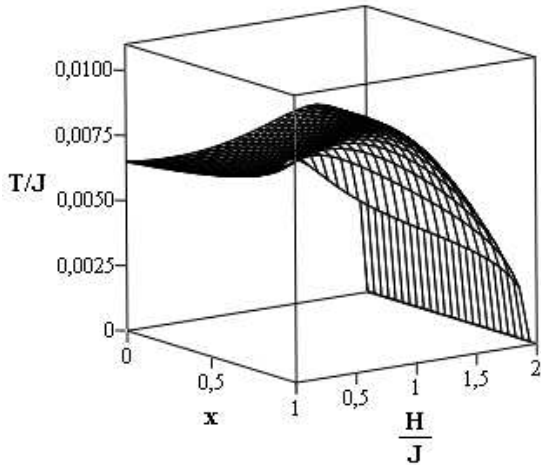}
\end{center}
\caption{The phase diagram for a quasi-1D spin-1/2 chain with spin
frustration, caused by nearest, NNN interactions and the ring
exchange. The N\'eel temperature is a function of the parameter $x$,
which shows how close the quantum critical point (caused by
spin-frustrating interactions) $x=1$ is, and the external magnetic
field.} \label{fig2}
\end{figure}
The maximum in the field dependence of the critical temperature, cf.
Fig.~1, is shifted towards low values of the field with the growth
of $x$, i.e. spin-frustrating NNN and multi-spin couplings can
reduce the domain of temperatures, at which re-entrant phase
transition can take place. We believe, that while the considered
model seems less realistic, the mentioned feature has the generic
nature for quasi-1D spin systems.

Phase diagrams, similar to the ones, presented in Figs.~1 and 2,
were obtained experimentally for real quasi-1D compounds \cite{faz}.
Phase $H-T$ diagrams in those compounds show maxima of the field
dependencies of critical temperatures. Namely, the ordering
temperature in studied quasi-1D copper oxides first increases with
the growth of the value of the field, reaches its maximum, and then
decreases to zero at the value of the field, where the spin chain
has the spin saturation. Notice, that in one of those compounds
measurements reveal different values of the ordering temperatures
for different directions of the external field, which can be caused
by the weak magnetic anisotropy of the intra-chain exchange
interactions.

\section{Conclusions}

In summary, we have used a simple analytical ansatz to calculate the
ordering temperature of a quasi-1D system, consisting of weakly
interacting quantum spin-1/2 chains with AF couplings in the
presence of the external magnetic field, when the weak inter-chain
coupling is taken into account in the mean field approximation, and
the characteristics of spin chains are obtained non-perturbatively.
Our results show that the field dependence of the critical exponents
for correlation functions of 1D subsystems plays a very important
role. In particular, that dependence determines the region of
possible re-entrant phase transition, governed by the field. We have
shown also how a quantum critical point between two phases of the 1D
subsystem, caused by spin-frustrating NNN and multi-spin ring-like
exchanges, affects the field dependence of the ordering temperature.
Our results qualitatively agree with the features, observed in
experiments on quasi-1D AF systems. We expect that our results are
generic for quasi-1D systems and that they can be helpful for
experimentalists, who study magnetic properties of such systems,
especially due to the recent progress in obtaining high values of
the magnetic field in experiments.

I thank the Deutsche Forschungsgemeinschaft for the financial
support via the Mercator Program. The support from the Institute of
Chemistry of the V.~Karazin Kharkov National University is
acknowledged.

\end{document}